\documentclass[twocolumn,superscriptaddress, preprintnumbers,amsmath,amssymb]{revtex4}

\usepackage{graphicx}
\usepackage{bm}
\include{epsf}
\usepackage{color}
\usepackage{xcolor}
\newcommand{\quotes}[1]{``#1''}
\usepackage[titletoc]{appendix}

\begin{document}


\title{Memory effects on epidemic evolution: The susceptible-infected-recovered epidemic model}

\author{M. Saeedian}
\affiliation{Department of Physics, Shahid Beheshti University, G.C., Evin, Tehran 19839, Iran}

\author{M. Khalighi}
\affiliation{Department of Physics, Shahid Beheshti University, G.C., Evin, Tehran 19839, Iran}

\author{N. Azimi-Tafreshi}
\affiliation{Physics Department, Institute for Advanced Studies in Basic Sciences, 45195-1159 Zanjan, Iran}

\author{G.~R. Jafari}
\affiliation{Department of Physics, Shahid Beheshti University, G.C., Evin, Tehran 19839, Iran}
\affiliation{School of Biological Sciences, Institute for Research in Fundamental Sciences (IPM), Tehran, Iran}
\affiliation{Center for Network Science, Central European University, H-1051, Budapest, Hungary}

\author{M. Ausloos}
\affiliation{GRAPES, rue de la Belle Jardini\`ere 483, B-4031, Angleur, Belgium}
\affiliation{ School of Management, University of Leicester, University Road, Leicester, LE1 7RH, United Kingdom}
\affiliation{ eHumanities group,  Royal Netherlands Academy of Arts and Sciences, Joan Muyskenweg 25, 1096 CJ, Amsterdam, The Netherlands}


  \date{\today}

\begin{abstract}
Memory has a great impact on the evolution of every process related to human societies. Among them, the evolution of an epidemic is directly related to the individuals' experiences. Indeed, any real epidemic process  is clearly sustained by a non-Markovian dynamics:  memory effects play an essential role in the spreading of diseases.
Including memory effects in the susceptible-infected-recovered (SIR) epidemic model seems very appropriate for such an investigation. Thus, the memory prone SIR  model dynamics is investigated using fractional derivatives.
The decay of long-range memory, taken as a power-law function, is directly controlled by the order of the fractional derivatives in the corresponding nonlinear fractional differential evolution equations.
Here we assume ``fully mixed'' approximation and show that the epidemic threshold is shifted to higher values than those for the memoryless system, depending on this  memory ``length'' decay exponent.
We also consider the SIR model on structured networks and study the effect of topology on threshold points in a non-Markovian dynamics. Furthermore, the lack of access to the  precise information about the initial conditions or the past events plays a very relevant role in the correct estimation or prediction of the epidemic evolution. Such a ``constraint'' is analyzed and discussed.

\end{abstract}

\pacs{87.23.Ge, 05.30.Pr, 05.70.Fh}

  \maketitle

\section{Introduction}
\label{s1}

The study of epidemiology, concerning the dynamical evolution of diseases within a population, has attracted much interest during the recent years \cite{review}. Mathematical models of infectious diseases have been developed in order  to integrate realistic aspects of disease spreading \cite{mathematicalmodels,models2,models3,models4}. A simple and commonly studied model, introduced by Kermack and McKendrick, is the susceptible-infected-recovered (SIR) model \cite{Kermack}. In this model, populations can be in each of three states: susceptible, infected and recovered (removed), denoted by S, I and R, respectively. Originally, it is assumed that susceptible individuals become  infected with a rate proportional to the fraction of infected individuals in the overall population (fully mixed approximation) and infected individuals recover at a constant rate. The epidemic process presents a (percolation) transition  between  a phase, in which the disease outbreak reaches a finite fraction of the population,  and a phase with only a limited number of infected individuals \cite{Grassberger,Stauffer}.
The model has also been investigated for population on lattices  (e.g. \cite{rhodes1996persistence,sander2003epidemic,tome2010critical}) or on networks (e.g. \cite{may2001infection,volz2007susceptible,parshani2010epidemic}).

For simplicity, we will keep the ``medical epidemic‌'' vocabulary hereafter. However, the model has also been interesting for describing nonmedical epidemics, such as for financial bubbles \cite{rotundo1,rotundo2}, migration \cite{ACSmigration}, opinion formation \cite{zhao2013sir,nizamani2014public} or internet ``worm propagation''  \cite{liljenstam2002mixed,kim2004measurement,mishra2010fuzzy}. SIR models with distributed delay and with discrete delay  have  also been studied \cite{beretta1995global,mccluskey2010complete}.

In the usual SIR model, it is assumed that all contacts transmit the disease with the same probability. Moreover, the transmission and recovery coefficients are constant. Hence the state of system at each time does not depend on the previous history of the system:  it is a memoryless, so-called Markovian, process.
However, real surveys show evidence of a non-Markovian spreading process  \cite{real,real2} in agreement with common expectation. The epidemic processes  evolution and control,  in human societies, cannot be considered without any memory  effect. When a disease spreads within a human population, the experience or knowledge of individuals about that disease should affect their response \cite{models4}. If people know about the history of a certain disease in the area where they live, they use different precautions, such as vaccination, when possible. Thus, some  endogenous controlled suppression of the spreading is expected, although other factors can help \cite{ecobichon1990chemical,wamai2008male,legreve2010preventing}.
However, knowledge about the history of a disease does not have the same influence at all times. Experience about the prevalence of a disease and precautions related to the ``old times'' are not always applicable or recommended, hence people tend to follow new strategies against the diseases. In other words,  memory of the earlier times could have less effect on the present  situation, as compared to more recent times. It can be expected that long-range memory effects decay in time more slowly than an exponential decay, but can typically behave like a power-law damping function.

While much effort has been made so far to determine exact epidemic thresholds in Markovian epidemic models \cite{threshold,threshold1,threshold2,thresholdNewmanPRE,stanley}, few works have been devoted to study the non-Markovian aspects of epidemic processes \cite{non-markov1,non-markov2}. Furthermore, in this work we focus on long-range memory effects, which means arbitrarily long history can be included. That is in contrast to short-term memory effects which have been extensively studied. For instance, Dodds and Watts \cite{Dodds} introduced a general model of contagion considering memory of past exposures to a contagious influence. The authors have argued that their model can fall into one of three universal classes, due to the behavior of fixed point curves. Also, in \cite{SIRSIR, Nature, SISSIS}, the authors consider ``implicit memory'' by applying asynchronous adapting in disease propagation. They show that this type of memory can lead to a first-order phase transition in outbreaks , thus hysteresis can arise in such  models \cite{SISSIS}.

It is here briefly recalled that fractional calculus is a valuable tool to observe the influence of  memory effects on the dynamics of systems \cite{Herrmann,West,Metzler,Hadis2}, and has been recently used in epidemiological models \cite{SIRfrac1,SIRfrac4,SIRfrac5,SIRfrac6}.
Typically, the evolution of epidemiological models is  described with  differential equations, the derivatives being of integer order. By replacing the ordinary time derivative by a fractional derivative, a time correlation function or memory kernel appears, thereby  making the state of the system dependent on all past states. Thus, it seems  that such a method based  on  derivatives with non-integer order, as introduced by Caputo for geophysics problems \cite{Caputo}, is  a very proper formalism for  such  non-Markovian  problems. Moreover, Caputo's formalism provides the advantage that  it is not
necessary to define the fractional order initial conditions, when solving such differential equations  \cite{Caputo,Podlubny,Podlubny1,fracappl}. Furthermore, the time correlation function, in the definition of Caputo fractional derivative,  is a power-law function,  which is flexible enough to reflect the fact that the contribution of more early states is noticeably less relevant than the contribution of  more recent ones on the present state of the dynamical system.

Most of the previous works have studied the epidemiological models with fractional order differential equations, from a
mathematical point of view. They mainly focused on presenting effective a mathematical methods in order to solve the corresponding differential equations \cite{SIRfrac4,Awawdeh,Young,AAM}. For instance in \cite{AAM} a mathematical tool (the multi step generalized differential transform method) is introduced to approximate the numerical solution of the SIR model with fractional differential equations. Also in \cite{SIRfrac5} the authors use fractional order differential equations for epidemic models and concentrate on the equilibrium
points of the models and their asymptotic stability of differential equations of fractional
order. Other variations of the SIR model with fractional derivatives have also been studied. For instance, Seo, \textit{et al}. introduced the SIR epidemic model with square root interaction of the susceptible and infected individuals and discussed the local stability analysis of the model \cite{Young}. Also in \cite{SIRfrac4}, numerical solution of the SIR epidemic model of fractional order with two levels of infection for the transmission of viruses in a computer network has been presented.

In all previous works, the authors rarely discuss the effect of fractional order
differential equations and memory on the epidemic thresholds and the macroscopic behavior of epidemic outbreaks. Hence, one question remains; we address  it in this paper:  How  does the system robustness change if memory is included in the SIR model? We also use the fractional differential equations, describing the SIR model on structured networks, to see the effect of topology on the evolution of the SIR model including memory effects.


Furthermore the lack of access to accurate information on  initial conditions  sometimes leads to  doubt about epidemic evolution predictions \cite{Shirazi}. The same type of difficulty  occurs in related problems, such as in opinion formation \cite{caram1,CaramCaifaAusloosProtoPRE}.  Moreover, it  may also happen in  certain cases  that individuals do not believe in old strategies in order to avoid the disease.
This means that the initial time for taking into account the  disease control memory is shifted toward more recent times:  thereafter,  the dynamics is evolving with a new fraction of susceptible and infected individuals, different from that predicted by the solution of the differential equations. In contrast, the fractional calculus  method allows us to choose any arbitrary initial time  at which the effect of initial conditions can be introduced on the spreading dynamics with a memory content.  The interest of fractional calculus will  appear  through such aspects in the core of the paper.

Thus, the paper is organized as follows. In Sec. \ref{s2}, following  Caputo's approach, we convert the differential equations of the standard SIR model to the fractional derivatives, thereby allowing us to consider memory effects. Using numerical analysis results (Sec. \ref{s3}), we discuss the influence of memory on the epidemic thresholds in Sec. \ref{s3finitetime}. We also discuss the dynamics of a non-Markovian epidemic process, when choosing different initial conditions or modifying the proportions of agents at a given time in Sec. \ref{s3initialconditions}. To complete our discussion, we study the dynamics of the model on structured networks in Sec.~\ref{s4}. We also point out that we have observed  qualitatively similar results for the SIS (susceptible-infected-susceptible) epidemic model. The conclusions are found in Sec.~\ref{s5}.
\section{memorial Process to Fractional equation}
\label{s2}

The evolution of the standard SIR model is described by a set of coupled ordinary differential equations for susceptible ($S$), infected ($I$), and recovered($R$) individuals, respectively given by,
\begin{eqnarray}
\label{SIR eq1}
\nonumber
\frac{dS(t)}{dt}&=& -\beta S(t)I(t),\nonumber \\
\frac{dI(t)}{dt}&=& \beta S(t)I(t)-\gamma I(t),\nonumber \\
\frac{dR(t)}{dt}&=& \gamma I(t),
\end{eqnarray}
in which, $\beta$ and $\gamma$ are infection and recovery coefficients, respectively.   The infected individual makes $\beta$  contacts per unit time producing new infections within a mean infectious time of order $1/\gamma$. The evolution of the model is controlled by quantity $\beta/\gamma$, such that above the epidemic threshold,   $(\beta/\gamma)_c$,  the disease spreads among a finite fraction of individuals.

These (ordinary) differential equations describe a Markov epidemic process, in which the state of individuals at each time step  does not depend on  previous steps.
The set of Eqs.~(\ref{SIR eq1}) can be solved  iteratively until time $t$. In particular, the fraction of susceptible individuals at time $t$, denoted as $S_t$, can be determined. In fact, $1-S_t$ is the size of outbreaks, i.e. the population that has or has had the disease until time $t$.

In order to observe the influence of  memory effects, first we rewrite the differential equations (\ref{SIR eq1}) in terms of time dependent integrals as follows,
\begin{eqnarray}
\label{SIR eq2}
\frac{dS(t)}{dt}&=&- \beta \int_{t_{0}}^{t} \kappa (t-t^{'})S(t^{'})I(t^{'})dt^{'},\nonumber\\
\frac{dI(t)}{dt}&=& \int_{t_{0}}^{t} \kappa (t-t^{'})\Big(\beta S(t^{'})I(t^{'})-\gamma I(t^{'})\Big)dt^{'},\nonumber\\
\frac{dR(t)}{dt}&=& \gamma \int_{t_{0}}^{t}\kappa (t-t^{'})I(t^{'})dt^{'},
\end{eqnarray}
in which, $\kappa(t-t^{'})$ plays  the role of  a time-dependent kernel and is equal to a delta function $\delta(t-t^{'})$ in a classical Markov process. In fact, any arbitrary function can be replaced by a sum of delta functions, thereby  leading to a  given type of time correlations. A proper choice, in order to include  long-term memory effects, can be a power-law function which exhibits a slow decay such that the state of the system at  quite early times also contributes to the evolution of the system. This  type of kernel guarantees the existence of scaling features as it is often intrinsic in most  natural phenomena.


Thus, let us consider  the following power-law correlation function  for $\kappa(t-t^{'})$:

\begin{eqnarray}
\label{SIR eq2a}
\kappa(t-t{'})=\frac{1}{\Gamma(\alpha -1) }(t-t^{'})^{\alpha-2},
\end{eqnarray}
in which $0<\alpha\leq1$ and $\Gamma(x)$ denotes the Gamma function. The choice of the coefficient $1/\Gamma{(\alpha-1)}$ and exponent $(\alpha-2)$ allows us to rewrite Eqs.~(2) to the form of fractional differential equations with the Caputo-type derivative.
If this kernel is  substituted into Eqs.~(\ref{SIR eq2}), the right hand side of the equations, by definition are fractional integrals of order $(\alpha-1)$ on the interval $[t_0,t]$, denoted by $_{t_0}\! \!D_t^{-(\alpha-1)}$. Applying a fractional Caputo derivative of order $\alpha-1$ on both sides of each Eq.~(\ref{SIR eq2}), and using the fact the Caputo fractional derivative and fractional integral are inverse operators, the following fractional differential equations can be obtained  for the SIR model:
\begin{eqnarray}
\label{SIR eq3}
_{t_0}^c\! \!D_t^{\alpha}S(t)&=&-\beta S(t)I(t),\nonumber\\
_{t_0}^c\! \!D_t^{\alpha}I(t)&=&\beta S(t)I(t)-\gamma I(t), \nonumber \\
_{t_0}^c\! \!D_t^{\alpha}R(t)&=&\gamma I(t),
\end{eqnarray}
where, $_{t_0}^c\! \!D_t^{\alpha}$ denotes the Caputo derivative of order $\alpha$, defined for an arbitrary function $y(t)$ as follows \cite{Caputo},
\begin{eqnarray}
_{t_0}^c\! \!D_t^{\alpha}~y(t)=\frac{1}{\Gamma(\alpha-1)}\int_{t_{0}}^{t}\frac{y'(\tau)d\tau}{(t-t_0)^\alpha}
\end{eqnarray}
Hence, the fractional derivatives,  when introducing a convolution integral with a power-law memory kernel, are useful to describe memory effects in dynamical systems. The decaying rate of the memory kernel (a time-correlation function) depends on $\alpha$. A lower value of $\alpha$ corresponds to more slowly-decaying time-correlation functions (long memory). Hence, in some sense, the strength (through the " length") of  the memory is controlled by $\alpha$. As $\alpha \rightarrow 1$, the influence of memory  decreases: the system tends toward a memoryless system.
Note that for simplicity, we assume  the same memory contributions (same value of $\alpha$) for different states of $S$, $I$ and $R$. Obviously, more complicated functions than Eq. (\ref{SIR eq2a}) and taking into account  different $\alpha_i$ ($i=1,2,3$)  could be investigated in further work to take into account different time scales.

Although analytical solutions of Eqs.~(\ref{SIR eq3}) are hard to obtain for the general case, they can be obtained at the early stage of the epidemic under a linearization approximation. In this case, it turns out that the number of infected individuals behaves as  a Mittag-Leffler function \cite{Podlubny}:
\begin{eqnarray}
\label{SIR eq4}
I(t)=E_{\alpha ,\zeta}(t) \equiv\sum_{k}  \frac{((\beta -\gamma ) t^{\alpha })^{k}}{\Gamma (\alpha k+\zeta )}
\end{eqnarray}
in which, $\zeta$ is a constant, - which depends on the initial conditions \cite{Podlubny}. In particular, for $\alpha=\zeta=1$, the Mittag-Leffler function is  the exponential function.
Thus, in the early stage of epidemic dynamics, the growth rate of the infected population in Eq.~(\ref{SIR eq4}) is positive, if $\beta-\gamma>0$. Therefore, the number of infected individuals grows exponentially in such a case, for $\beta>\gamma$, as  of course  it is expected for the standard memoryless SIR model. The same  reasoning applies in order to determine  the epidemic threshold  for $\alpha<1$.

\section{Numerical results} \label{s3}
Let it be reemphasized that Eqs.~(\ref{SIR eq3}) consist in a system of coupled non-linear differential equations of fractional order, in the following general form
\begin{eqnarray}
\label{SIR eq5}
_{t_0}^c\! \!D_t^{\alpha}y^{(i)}(t)&=& f^{(i)}(t,y^{(1)}(t),y^{(2)}(t),y^{(3)}(t))\nonumber\\
y^{(i)}(t_0)&=&y^{(i0)},
\end{eqnarray}
where, $i=1,2,3$ and $y^{(1)},y^{(2)},y^{(3)}$ denote $S, I, R$ cases respectively. Also, $y^{(i0)}$ are constants which indicate the initial conditions.

To solve the equations, we use the the predictor corrector
algorithm, which is well known for obtaining a numerical solution of  first order problems \cite{Diethelm,Diethelmer,Garrappa}.
It is assumed that there exits a unique solution for each of $y_{(i)}$ on the interval $[0,T]$ for a given set of initial conditions.
Considering  a uniform grid $\{ t_{n} = nh : n=0,1,2,...,N\}$, in which $N$ is an integer and $ h\equiv T/N$, each Eq.~(\ref{SIR eq5}) can be rewritten in a discrete form,
\begin{eqnarray}
\label{SIR eq6}
y^{(i)}_n=y^{(i)}_0+ h^{\alpha}\sum_{k=0}^{n-1}b_{n-k-1}f^{(i)}_k,
\end{eqnarray}
where the coefficients $b_{n-k-1}$  refer to the contribution of each of the $n-1$ past states on the present state of $n$ . The coefficients  are given by
\begin{eqnarray}
\label{SIR eq7}
b_{n-k-1}=\frac{(n-1-k)^{\alpha}-(n-k)^{\alpha}}{\Gamma{(\alpha+1)}}
\end{eqnarray}

Thereby after solving Eq.~(\ref{SIR eq5}), numerically, the influence of memory  on the evolution of the SIR epidemic model can be analyzed. As mentioned in the Introduction, let us consider two pertinent aspects successively: the finite time behavior and the role of changing initial conditions.
%
\begin{figure}[t]
\begin{center}
\scalebox{0.42}{\includegraphics[angle=0]{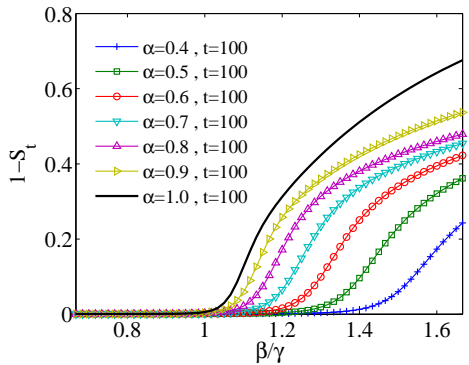}}
\end{center}
\caption{Outbreak size $1-S_{t}$ for a SIR system  having evolved until time $t=100$, vs. the parameter defining the threshold: $\beta/\gamma$, when including memory effects. Each curve corresponds to a  different value of $\alpha$, as indicated in the inset.  As $\alpha$ decreases, the epidemic threshold $(\beta/\gamma)_c$ shifts to  higher values. }
\label{f1}
\end{figure}
\subsection{Epidemic threshold  at finite times}\label{s3finitetime}
Let us compare the evolution of a system  including memory effects with the memoryless case. We solve Eq.~(\ref{SIR eq5}) with initial conditions $y^{(10)}=S_0=1-\epsilon$, $y^{(20)}=I_0=\epsilon$. Fig.~\ref{f1} shows the size of  the outbreak for different values of $\alpha$, measured  util  $t=100$ and for $\epsilon=10^{-4}$. The size of outbreak,
$1-S_t$, is zero (with accuracy $10^{-4}$ ) for small values of $\beta/ \gamma$. The specific value of $\beta/ \gamma$, in which the epidemic size starts to get a non-zero value, is identified as the epidemic threshold point.

\begin{figure}[t]
\begin{center}
\scalebox{0.34}{\includegraphics[angle=0]{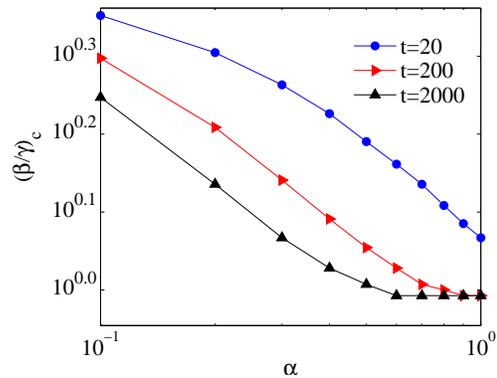}}\\
\end{center}
\caption{Variation of threshold point vs. $\alpha$ for different finite times $t=20,200,2000$. For each time, the epidemic threshold is shifted to higher values with decreasing $\alpha$. The axes are logarithmic and the numbers are presented as 10-base exponential notation.
}
\label{f1a}
\end{figure}

\begin{figure}[t]
\begin{center}
\scalebox{0.42}{\includegraphics[angle=0]{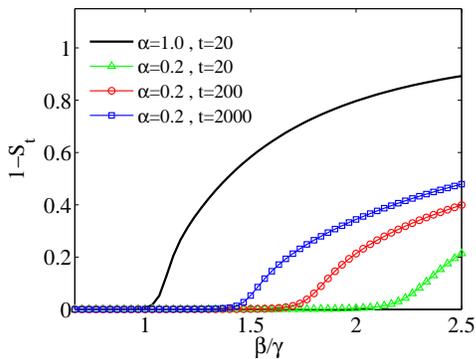}}
\end{center}
\caption{Order parameter $1-S_{t}$ for a SIR system having evolved until  time $t$, when  including much memory  ($\alpha=0.2$). Each curve corresponds to a different finite time $t$, as indicated in the inset. The threshold  values can be compared with that of  the  corresponding epidemic threshold for a memoryless system, i.e. when $\alpha=1$ (and $t =20$). The curves for  $\alpha = 1$ at $t =200$ and $t = 2000$ are not drawn for better readability.}
\label{f2}
\end{figure}

\begin{figure}[t]
\begin{center}
\scalebox{0.34}{\includegraphics[angle=0]{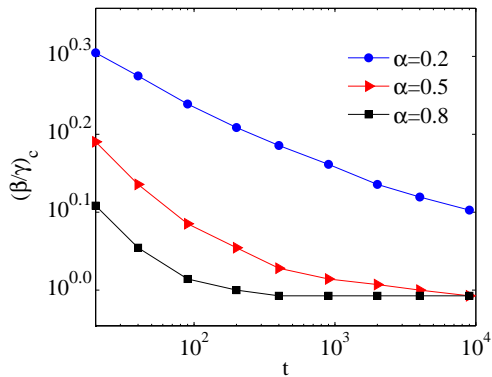}}\\
\end{center}
\caption{Variation of threshold point vs. $t$ for different values of $\alpha=0.2,0.5,0.8$. For each $\alpha$, the epidemic threshold is shifted to lower values with increasing finite time. The axes are logarithmic and the numbers are presented as base 10 exponential notation.
}
\label{f2a}
\end{figure}

The stationary time for a memoryless system ($\alpha=1$) is $t=100$. With decreasing the value of $\alpha$ (including memory) the system needs much time to reach the stationary state. Hence at $t=100$, the threshold point is shifted to the higher value of $\beta/\gamma$.
Figure.~\ref{f1a} shows that the threshold point is increased with decreasing of $\alpha$ for a finite time $t$. Furthermore, as  it  can be seen in Fig.~\ref{f1}, the size of outbreaks decreases for decreasing $\alpha$.
Let the interval $[t_{0}, t]$ be  the time interval in which memory effects are taken into account.
In Fig.~\ref{f2}, we compare the evolution of the model with memory for different values of  the finite time $t$. The memory effects are considered  for a weight $\alpha=0.2$. It is seen that as time evolves  the influence of memory    decreases, since memory effects decay in time like a power-law function. Hence, the epidemic threshold  shifts to lower values of effective infection rate $\beta/ \gamma$ and approaches  the threshold of the memoryless model ($\alpha=1$). The curves for  $\alpha = 1$ at $t =200$ and $t = 2000$ are hardly distinguishable from the curve at $t = 20$ and are not drawn for better readability. The variation of threshold point, with increasing finite time, is shown in Fig.~\ref{f2a}. Furthermore, for a given  $\beta/ \gamma$ value, it appears that there is more time available for  disease spreading,  whence more individuals become infected.

\subsection{Initial conditions}\label{s3initialconditions}

Recall that the dynamics of a non-Markovian process is directly influenced by all events from the beginning of the process. However,  some loss of information about some period of time in the past  may lead one to consider  that the influence of memory  might  not  need to be considered as continuous. It may happen, in many social networks, that  individuals do not have enough information about the history of a disease, as recent cases  and studies indicate; e.g. see \cite{Eichelberger,BKJohns,morens2015forgotten,tomori2015will}. Only after several individuals have already been infected, do people start to increase their knowledge about the disease and  take different precautions. The question arises on how the ``initial time''  at  which a non-Markovian process is started, affects  the subsequent dynamics of the process.
\begin{figure}[t]
\begin{center}
\scalebox{0.42}{\includegraphics[angle=0]{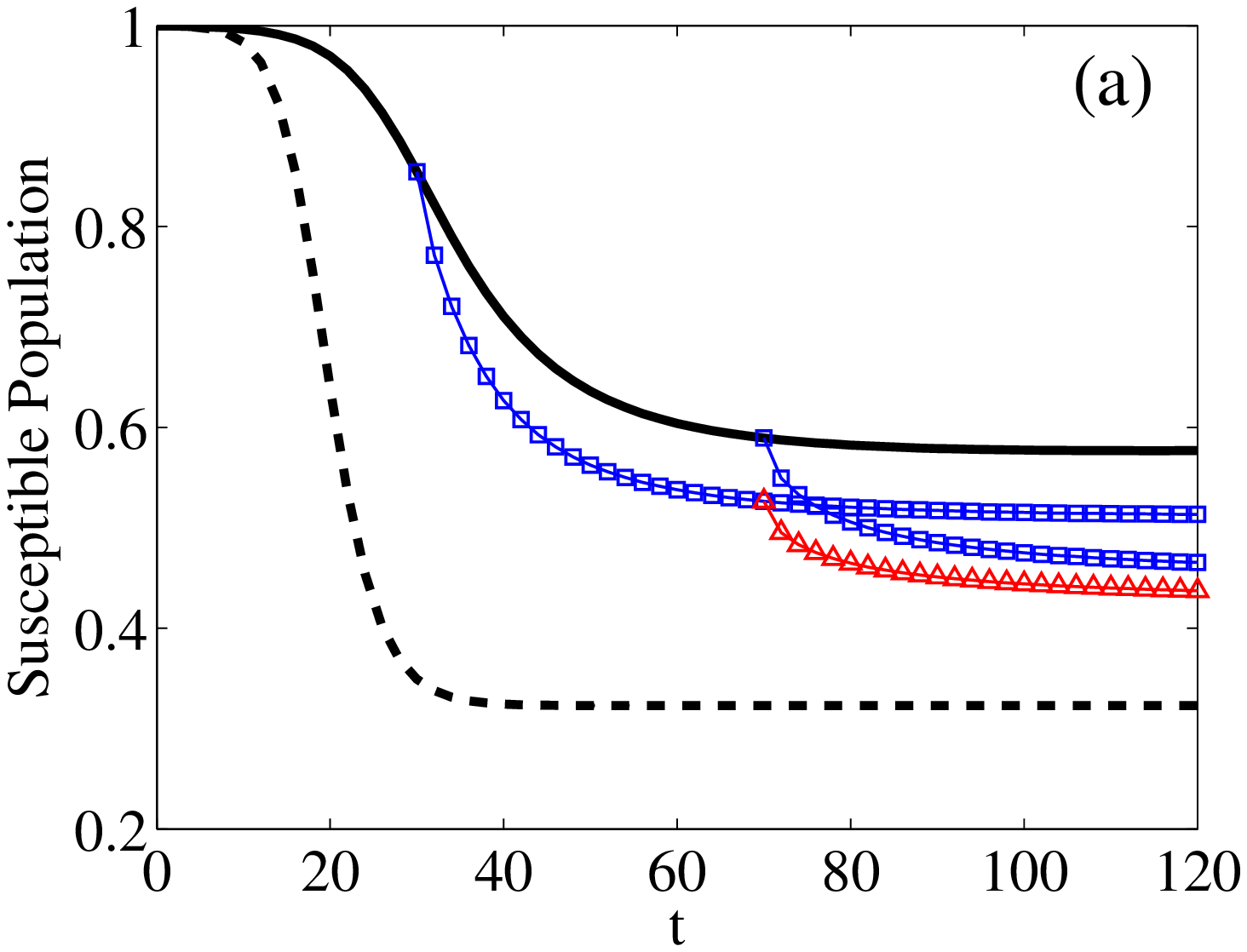}}\\
\scalebox{0.42}{\includegraphics[angle=0]{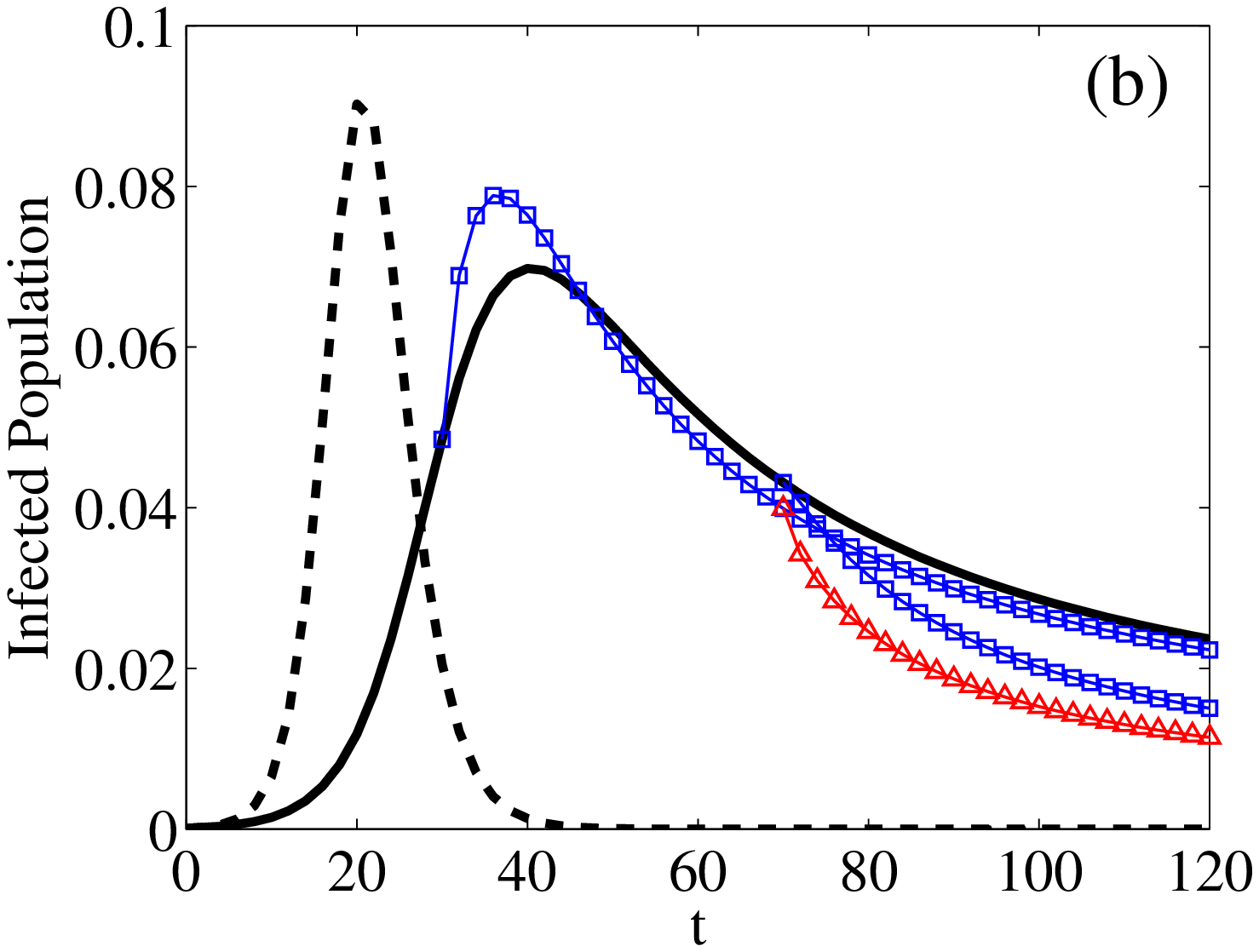}}\\
\scalebox{0.42}{\includegraphics[angle=0]{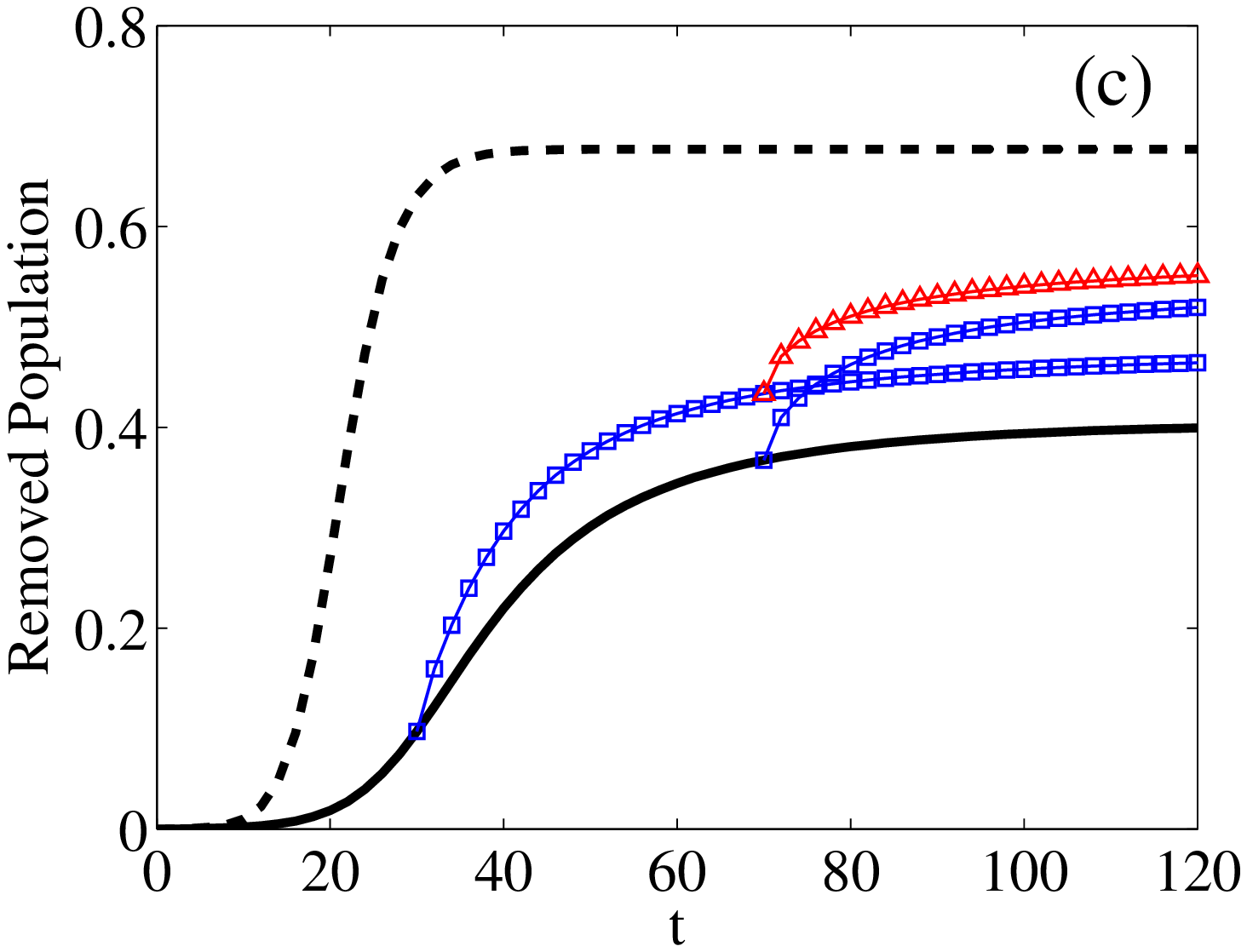}}\\
\end{center}
\caption{Effect of different initial times on the dynamics of a non-Markovian process. The curves denote the fraction of $(a)$ susceptible, $(b)$ infected and $(c)$ removed individuals. Dashed and solid lines correspond to Markovian and non-Markovian processes, respectively, started from $t=0$. The curves with symbols, correspond to the dynamics of non-Markovian processes, started from non-zero initial times with different initial conditions.
}
\label{f3}
\end{figure}

If two Markovian processes  start at  two different times, the evolution of both processes  is identical. However, the scenario is  quite different for a non-Markovian process, i.e. in which the memory  plays a  role.  This is illustrated through Fig.~\ref{f3} where  the fractions of susceptible, infected, and removed individuals are compared in the case of two Markov and non-Markov epidemic processes. Continuous and dashed (black) lines correspond to a system with and without including memory effects, respectively, evolving from the same initial time $t=0$. As it can be seen the fraction of susceptible individuals is greater in a system with inclusion of memory effects  with respect  to that   ignoring the memory (Fig.~\ref{f3}(a)). In other words, the experience and knowledge  which individuals have about the disease are obviously  helping them to protect themselves against the disease. Equivalently, in a system including the memory effects, the infection grows more slowly  as  seen in Fig.~\ref{f3}(b).

Thereafter, consider that a non-Markovian process, including memory effects, has  evolved until a specific time $t_1$.
Let  the process be continuing its  evolution, but let the memory of  the system be removed at that time. This corresponds to having a new initial time and new  initial conditions for the epidemics spreading.   The process can be continued without or with memory.  The Markovian case is trivial thereafter and thus not discussed.  Instead, consider that  memory effects are only taken  into account  at this starting  ``new initial time''.  In other words, let the population  ignore the disease control history (memory)   until   $t_1$; let  the system continue its dynamics but taking into account  memory effects  thereafter from $t_1$. The  initial conditions for the evolution of the system are  now a fraction of susceptible and infected individuals at time $t_1$. The curves with square symbols in Fig.~\ref{f3}, correspond to  what happens for different  ``new initial times'' $t_1=30,70$, for the dynamics of  such a non-Markovian epidemic process. As it can be seen, at the beginning of the dynamics, the fraction of susceptible individuals is reduced, since people do not know about the disease. However,  as  soon as it is influenced by memory,  the system  becomes more resilient to the spreading. Hence, the fraction  of  $S$    individuals remains greater  as compared to that with a memoryless system, having started  at $t=0$. In a similar manner, the fraction of infected and removed individuals deviate from the original one and tend toward the populated states of a memoryless system when  the memory from further past times is included. In this case,
the curves become closer to the dashed curve corresponding to  a memoryless system. That means that the system loses the information related to past times and   tends to present  a behavior  similar  to a  memoryless system.

Finally, one can  consider ``to remove the memory'' of an epidemic process  at various times. At each time step, the system  is supposed to lose (or practically negate)  the  information  about  the disease before  some ``re-awareness time"  (see also \cite{funk2009spread}) and to continue  its dynamics regardless of  the past.  For illustration, consider  the case of such a sudden  awareness and its impact on epidemic outbreaks case  through  Fig.~\ref{f3};  the system loses its memory at times $t_1=30$ and $t_2=70$, i.e.  the dynamics is stopped at $t_1=30$, then  is  continued until $t_2=70$, removing all the history of the system before that time, next reintroducing the memory dynamics again at $t_2=70$:
see the (red curves with) triangular symbols in Fig.~\ref{f3} corresponding to   this case    of  a double  ``loss of memory.''

Notice that in this  particular illustrative case, the behavior of  the system is  seen to be close  to the dynamics of a memoryless system, since contributions of the memory of the system are  sometimes removed. Such an illustration points to the interest of the model in order to compare it with the case of epidemics spreading waves \cite{BKJohns}; for completeness, let it be pointed out that the connection of periodic epidemics  to SIR models has been already mentioned \cite{Greenhalgh1988}: flu is yearly recurrent.  Notice also that  the value of  $\alpha$ could be modified at each new awareness time, but this investigation goes outside the present  paper.
\section{The model on structured networks}
\label{s4}

So far we have considered the fully mixed approximation,
such that an infected individual is equally likely to
spread the disease to any other individual. However, in the real world an individual connects to a small
fraction of people. Hence, as is well known, more realistic modeling can be studied through networks,
where their topology has a significant effect on the epidemic process \cite{Moore,Kuperman,SFepidemic}.
For homogeneous networks, each individual has the same number of connections $k\approx  \langle k\rangle$
and disease propagates with spreading rate $\beta \langle k\rangle$. In this case, it is obvious that the
epidemic threshold $(\frac{\beta}{\gamma})_c$ is simply replaced by $(\frac{\beta}{\gamma})_c\langle k\rangle$. It is also true for the case of fractional differential Eqs.~4. In other words, threshold point in Fig.~\ref{f1} for each value of $\alpha$ is shifted to $(\frac{\beta}{\gamma})_c\langle k\rangle$.

Now, let us consider heterogeneous scale free networks with degree distribution $P(k)\sim k^{-\lambda}$.
In heterogeneous mean field approximation, it is assumed that all nodes are statistically equivalent and thus one can consider
groups of nodes with the same degree $k$. With this assumption the ordinary differential equations describing the SIR model are given by
\begin{eqnarray}
\label{SIRN eq1}
\nonumber
\frac{ds_{k}(t)}{dt}&=& -\beta ks_{k}(t)\Theta_{k}(t),\nonumber \\
\frac{di_{k}(t)}{dt}&=& \beta ks_{k}(t)\Theta_{k}(t)-\gamma i_{k}(t),\nonumber \\
\frac{dr_{k}(t)}{dt}&=& \gamma i_{k}(t),
\end{eqnarray}
where
\begin{eqnarray}
\Theta_k(t)=\frac{\sum_{k-1}kP(k)i_{k}(t)}{\sum_{k}kP(k)}.
\end{eqnarray}
and $i_k, s_k$ and $r_k$ denote the density of infected, susceptible and removed nodes in each group, respectively.
It was turned out that in scale-free networks characterized by a degree exponent $2<\lambda \leq3$, there is no epidemic threshold \cite{SFepidemic}.

\begin{figure}[t]
\begin{center}
\scalebox{0.42}{\includegraphics[angle=0]{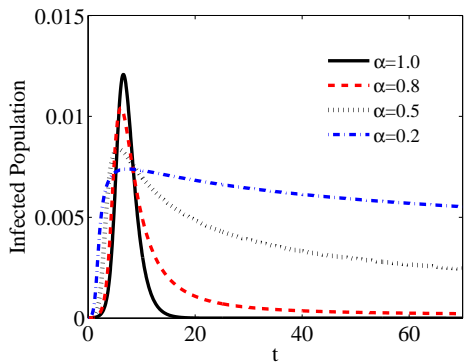}}\\
\end{center}
\caption{Fraction of infected individuals versus time for the SIR model on a scale-free network with degree exponent $\lambda=3$ and for different values of $\alpha$.
}
\label{f6}
\end{figure}

\begin{figure}[t]
\begin{center}
\scalebox{0.32}{\includegraphics[angle=0]{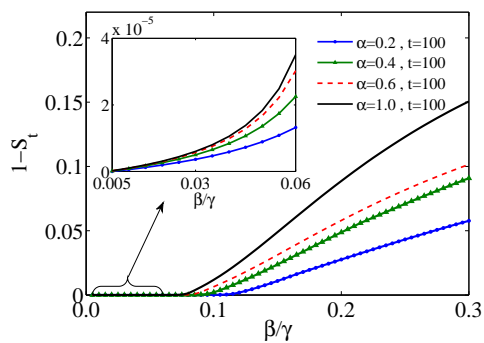}}\\
\end{center}
\caption{Outbreak size, $1-S_{t}$, for the SIR model on a scale-free network with degree exponent $\lambda=3$  in terms of $\beta/\gamma$. The dynamics is evolved until time $t=100$, when including memory effects. Each curve corresponds to a  different value of $\alpha$, as indicated in the insert.
}
\label{f7}
\end{figure}

Following the same procedure, presented in Sec.\ref{s2}, we can rewrite Eqs.~\ref{SIRN eq1}
to the fractional derivatives, as follows
\begin{eqnarray}
\label{SIRsf}
_{t_0}^c\! \!D_t^{\alpha}s_{k}(t)&=&-\beta s_{k}(t)\Theta_{k}(t),\nonumber\\
_{t_0}^c\! \!D_t^{\alpha}i_{k}(t)&=&\beta s_{k}(t)\Theta_{k}(t)-\gamma i_{k}(t), \nonumber \\
_{t_0}^c\! \!D_t^{\alpha}r_{k}(t)&=&\gamma i_{k}(t),
\end{eqnarray}
For a network with degree exponent $\lambda=3$, we solve Eqs.~(\ref{SIRsf}) numerically. Figure~\ref{f6} shows the evolution of the fraction of total infected individuals $i(t)=\sum_k P(k)i_k(t)$, with considering memory effects with different values of $\alpha$. While for a memoryless SIR model $(\alpha=1)$, the system reaches a stationary state after a short time ($t\simeq 20$), the stationary time is increased with decreasing the value of $\alpha$. Furthermore, we obtain the size of outbreaks at a finite time. Figure~\ref{f7} shows $1-S_t$, measured with accuracy $10^{-5}$ until $t=100$ for different values of $\alpha$. As we can see the epidemic threshold is always zero, as it is for Markov epidemic spreading on scale-free networks with $\lambda=3$. However the size of epidemic decreases with decreasing $\alpha$. The same results are obtained for networks with $2<\lambda<3$. However for $\lambda>3$, the epidemic threshold is shifted with including the memory, similar to what is observed for the homogenous networks.
\section{Conclusion}
\label{s5}
Memory plays a significant role in the evolution of many real dynamical processes, including the  cases of epidemic spreading. Here we have reported a study on the evolution of the SIR epidemic model, considering memory effects. Using the fractional calculus technique, we show  that the dynamics of  such a system depends on the strength of memory effects, controlled by the order of fractional derivatives $\alpha$. At finite times, including memory effects, the epidemic threshold  $(\beta/\gamma)_c$ is shifted to higher values than those for memoryless systems, at values   depending on the memory decay rate  $\alpha$. In the case that the model evolves on heterogeneous scale-free networks with $2<\lambda\leq3$, the threshold point is always zero. However, the fraction of individuals who are infected or recovered, is reduced if the  memory ``length" increases. Hence, memory  renders the system more robust against the disease spreading. If the epidemic process evolves  further in  time,  for a fixed  memory strength,  (i)  the disease can infect more individuals and (ii) the epidemic threshold is shifted to smaller values and tends to the
memoryless case values.

Furthermore, we have shown the following result: the evolution of an epidemic process, including memory effects, much depends on the fraction of infected individuals at the beginning of the memory effect insertion in the evolution. During a non-Markovian epidemic process, if the system abruptly loses its memory at a definite time and  if from that time on, one lets the non-Markovian process continue again,  starting with the number of infected individuals at that time, the dynamics of the system deviates from the basic case, in which the system continuously includes memory effects  from the beginning of the process.

Our observations are obtained from a simple epidemiological model: the SIR model. Obviously many parameters are  here  assumed  to be constant. We are aware that some, e.g. policy, feedback might influence the parameter values.  They may depend on space, groups, and time.  External field conditions may  also surely influence real aspects. However, we guess that many qualitative behaviors as those presented  here are  likely to be quite generally found in reality.  More advanced epidemic models, based on  various types of complex networks  are surely  interesting subjects for further investigations, in line with investigations  such as, e.g.,  in \cite{threshold,threshold1,threshold2,thresholdNewmanPRE,stanley}.  We also wish to point out that we have observed qualitatively similar results for the SIS epidemic model.
Finally, we  may claim that our results are not limited to the epidemiological  (``medical'') models but also can be extended for analogous epidemic spreading of  rumors, gossip, opinions,  religions,  and other topics pertinent to  epidemics on many social networks.

\appendix

\section{}
\begin{figure}[t]
\begin{center}
\scalebox{0.36}{\includegraphics[angle=0]{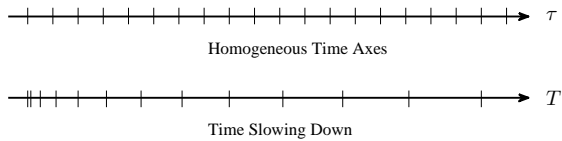}}
\end{center}
\caption{Schematic comparison between homogeneous and fractional time axes. }
\label{Af1}
\end{figure}

It could be instructive to study fractional order operators within a geometric interpretation (see interesting references in \cite{Podlubny1}). Here we compare the time scales in fractional- and integer-order dynamics.
To image a geometric interpretation, let us consider the right-sided fractional integral of order $\alpha$,
\begin{eqnarray}
\label{xfrac1}
x_{t}^{\alpha}=\frac{1}{\Gamma(\alpha)}\int_{0}^{t}v(\tau)(t-\tau)^{\alpha-1}d\tau
\end{eqnarray}
and write it in the form
\begin{eqnarray}
\label{xfrac2}
x_{t}^{\alpha}=\int_{0}^{t}v(\tau)dT_t({\tau})
\end{eqnarray}
where
\begin{eqnarray}
\label{xfrac3}
T_t({\tau})=\frac{1}{\Gamma(\alpha+1)}\{t^{\alpha}-(\tau-t)^{\alpha}\}
\end{eqnarray}
If we compare Eq.~\ref{xfrac2} with its counterpart $x_{t}^{\alpha}=\int_{0}^{t}v(\tau)d\tau$, in the homogeneous time scheme, the main difference is related to the different time variables, $T$ and $\tau$.
Notice that time variable $T_t(\tau)$ has a scaling property. If we take $t_1=kt$ and $\tau_1=k\tau$, then $T_{t_1}(\tau_1)=k^{\alpha}T_t(\tau)$. Hence, in the fractional order dynamics, the time is \quotes{accelerating} in the early time and after that it is \quotes{slowing down}, as sketched in Fig.~\ref{Af1}.
In this case, the \quotes{passing time} in the two axes of time is not the same. For this reason, in epidemic ``fractional'' dynamics, the epidemic threshold is shifted to the higher values. A lower $\alpha$ indicates a \quotes{stronger} (long lasting) memory and a more pronounced shift of threshold point. However, if one waits long enough, the same behavior will be observed in both fractional and usual homogeneous time. In fractional dynamics, after a \quotes{very long} time, the threshold point coincides with the one appearing in integer-order dynamics.
%

\end{document}